\documentclass[useAMS]{mn2e}
 \usepackage{times}
 \usepackage{graphicx}
 \usepackage{epsfig}
 \usepackage{amssymb}

 \title[Yet another Earth quasi-satellite]
       {Asteroid 2014~OL$_{339}$: yet another Earth quasi-satellite}

 \author[C. de la Fuente Marcos and R. de la Fuente Marcos]
        {C.~de~la~Fuente~Marcos\thanks{E-mail: nbplanet@fis.ucm.es}
         and
         R. de la Fuente Marcos \\
         Universidad Complutense de Madrid,
         Ciudad Universitaria, E-28040 Madrid, Spain}
 \date{Accepted 2014 September 19.
       Received 2014 September 15;
       in original form 2014 September 5}
 \pubyear{2014}
 
\begin{document}
  \maketitle

  \begin{abstract}
     Our planet has one permanently bound satellite --the Moon--, a likely 
     large number of mini-moons or transient irregular natural satellites, 
     and three temporary natural retrograde satellites or quasi-satellites. 
     These quasi-moons --(164207) 2004~GU$_{9}$, (277810) 2006~FV$_{35}$ and 
     2013~LX$_{28}$-- are unbound companions to the Earth. The orbital 
     evolution of quasi-satellites may transform them into temporarily bound 
     satellites of our planet. Here, we study the dynamical evolution of the 
     recently discovered Aten asteroid 2014~OL$_{339}$ to show that it is 
     currently following a quasi-satellite orbit with respect to the Earth. 
     This episode started at least about 775 yr ago and it will end 165 yr 
     from now. The orbit of this object is quite chaotic and together with 
     164207 are the most unstable of the known Earth quasi-satellites. This 
     group of minor bodies is, dynamically speaking, very heterogeneous but 
     three of them exhibit Kozai-like dynamics: the argument of perihelion of 
     164207 oscillates around -90\degr, the one of 277810 librates around 
     180\degr and that of 2013~LX$_{28}$ remains around 0\degr. Asteroid 
     2014~OL$_{339}$ is not currently engaged in any Kozai-like dynamics. 
  \end{abstract}


  \section{Introduction}
     The term "quasi-satellite" was first used in a scientific publication by Danielsson \& Ip (1972) while trying to explain the resonant
     behaviour of the near-Earth Object (NEO) 1685 Toro (1948 OA). However, this early mention was not directly connected with its current 
     use. It is now generally accepted that the term was first introduced and popularized among the scientific community by Mikkola \& 
     Innanen (1997), although the concept behind it was initially studied by Jackson (1913) and the energy balance associated with the 
     resonant state was first explored by H\'enon (1969), who coined the term "retrograde" satellites to refer to them. Further analyses 
     were carried out by Szebehely (1967), Broucke (1968), Benest (1976, 1977), Dermott \& Murray (1981), Kogan (1989) and Lidov \& 
     Vashkov'yak (1993, 1994a,b). Most of this early work was completed within the framework of the restricted elliptic three-body problem. 
     The quasi-satellite dynamical state is a specific configuration of the 1:1 mean motion resonance with a host planet in which the object 
     involved appears to travel around the planet but is not gravitationally bound to it, i.e. the body librates around the longitude of its 
     associated planet but its trajectory is not closed. 

     The first minor body to be confirmed to pursue a quasi-satellite orbit was 2002~VE$_{68}$ that is companion to Venus (Mikkola et al. 
     2004). Objects in this dynamical state have been found following Ceres and Vesta (Christou 2000b; Christou \& Wiegert 2012), Jupiter 
     (Kinoshita \& Nakai 2007; Wajer \& Kr\'olikowska 2012), Saturn (Gallardo 2006), Neptune (de la Fuente Marcos \& de la Fuente Marcos 
     2012c) and Pluto (de la Fuente Marcos \& de la Fuente Marcos 2012a). So far, Jupiter has the largest number of documented 
     quasi-satellites with at least six, including asteroids and comets (Wajer \& Kr\'olikowska 2012). Our planet comes in second place with 
     three detected quasi-satellite companions: (164207) 2004~GU$_{9}$ (Connors et al. 2004; Mikkola et al. 2006; Wajer 2010), (277810) 
     2006~FV$_{35}$ (Wiegert et al. 2008; Wajer 2010) and 2013~LX$_{28}$ (Connors 2014). As such, these objects are not real, 
     gravitationally bound satellites but, from Earth's point of view, they appear to travel in the retrograde direction around it over the 
     course of a year although they actually orbit (in the prograde direction) the Sun. Large amounts of interplanetary dust particles are 
     also temporarily trapped in Earth's quasi-satellite resonance (Kortenkamp 2013) and our planet hosts a small population of transient 
     irregular natural satellites or mini-moons that may have been quasi-satellites before becoming temporarily bound to the Earth (Granvik, 
     Vaubaillon \& Jedicke 2012; Bolin et al. 2014)

     Here, we show that the recently discovered asteroid 2014~OL$_{339}$ is a quasi-satellite companion to the Earth. The object was 
     originally selected as a co-orbital candidate because of its small relative semimajor axis, $|a - a_{\rm Earth}| \sim$ 0.0002 au; 
     $N$-body calculations are used to confirm its current quasi-satellite engagement with our planet. This paper is organized as follows.
     In Section 2, we briefly outline our numerical model. Section 3 focuses on 2014~OL$_{339}$. Section 4 reviews the current dynamical
     status of 164207, 277810 and 2013~LX$_{28}$, using their latest orbital solutions. Section 5 provides a comparative dynamical analysis 
     between 2014~OL$_{339}$ and the other three Earth quasi-satellites. Our conclusions are summarized in Section 6.

  \section{Numerical model}
     Here, we use $N$-body calculations to study the librational properties of the principal resonant angle of 2014~OL$_{339}$ with the 
     Earth in order to understand its current dynamical status. As an Earth co-orbital candidate, the key object of study is the oscillation 
     of the difference between the mean longitudes of the object and the Earth or relative mean longitude, $\lambda_{\rm r}$. The mean 
     longitude of an object is given by $\lambda$ = $M$ + $\Omega$ + $\omega$, where $M$ is the mean anomaly, $\Omega$ is the longitude of 
     the ascending node and $\omega$ is the argument of perihelion (see e.g. Murray \& Dermott 1999). An object is co-orbital to the Earth 
     if $\lambda_{\rm r}$ oscillates (librates) around a constant value; if $\lambda_{\rm r}$ can take any value (circulates), then we have 
     a passing object. If $\lambda_{\rm r}$ librates around 0$^{\circ}$, we have the quasi-satellite state; the minor planet orbits the Sun 
     in an approximate ellipse with the same (mean) period as the Earth. However, when viewed in a frame of reference that corotates with 
     the Earth, the quasi-satellite follows a retrograde path around our planet over the course of an orbital period, the sidereal year. In 
     principle, such motion is stabilized by the host planet. The stability of quasi-satellite orbits has been studied by Mikkola et al. 
     (2006) and Sidorenko et al. (2014). 

     The numerical simulations presented here were completed using a Hermite integration scheme (Makino 1991; Aarseth 2003). The standard 
     version of this direct $N$-body code is publicly available from the IoA web site\footnote{http://www.ast.cam.ac.uk/$\sim$sverre/web/pages/nbody.htm}. 
     Our model Solar system includes the perturbations by the eight major planets and treats the Earth and the Moon as two separate objects, 
     it also incorporates the barycentre of the dwarf planet Pluto--Charon system and the ten most massive asteroids of the main belt, 
     namely, (1) Ceres, (2) Pallas, (4) Vesta, (10) Hygiea, (31) Euphrosyne, 704 Interamnia (1910 KU), 511 Davida (1903 LU), 532 Herculina 
     (1904 NY), (15) Eunomia and (3) Juno. Relative errors in the total energy at the end of the simulations are $< 1 \times 10^{-15}$. The 
     equivalent error in the total angular momentum is several orders of magnitude smaller. Additional details can be found in de la Fuente 
     Marcos \& de la Fuente Marcos (2012b) which also discusses 2002~VE$_{68}$, the first documented quasi-satellite. 
%
%
     \begin{table*}
      \fontsize{8}{11pt}\selectfont
      \tabcolsep 0.12truecm
      \caption{Heliocentric Keplerian orbital elements of asteroids 2014~OL$_{339}$, (164207) 2004~GU$_{9}$, (277810) 2006~FV$_{35}$ and 
               2013~LX$_{28}$, all current quasi-satellites of our planet. Values include the 1$\sigma$ uncertainty. The orbit of 
               2014~OL$_{339}$ is based on 27 observations with a data-arc span of 36 d. The orbits are computed at Epoch JD 245\,7000.5 
               that corresponds to 0:00 \textsc{ut} on 2014 December 9 (J2000.0 ecliptic and equinox. Source: JPL Small-Body Database. Data 
               as of 2014 September 15.)
              }
      \begin{tabular}{cccccc}
       \hline
                                                               &   &   2014~OL$_{339}$             &   2004~GU$_{9}$             
                                                                   &   2006~FV$_{35}$              &   2013~LX$_{28}$                \\
       \hline
        Semimajor axis, $a$ (au)                               & = &   0.999 388$\pm$0.000 007     &   1.001 268128$\pm$0.000 000003 
                                                                   &   1.001 27736$\pm$0.000 00002 &   1.001 5884$\pm$0.000 0012     \\
        Eccentricity, $e$                                      & = &   0.460 67  $\pm$0.000 03     &   0.136 2649$\pm$0.000 0006   
                                                                   &   0.377 531$\pm$0.000 005     &   0.452 052$\pm$0.000 011       \\
        Inclination, $i$ ($^{\circ}$)                          & = &  10.190 65 $\pm$0.000 5       &  13.648 35$\pm$0.000 05       
                                                                   &   7.101 62$\pm$0.000 13       &  49.976 1$\pm$0.000 3           \\
        Longitude of the ascending node, $\Omega$ ($^{\circ}$) & = & 252.223 2$\pm$0.001 1         &  38.675 83$\pm$0.000 03       
                                                                   & 179.541 89$\pm$0.000 08       &  76.681 00$\pm$0.000 02         \\
        Argument of perihelion, $\omega$ ($^{\circ}$)          & = & 289.656 4$\pm$0.000 5         & 280.332 87$\pm$0.000 05       
                                                                   & 170.845 8$\pm$0.000 2         & 345.781 8$\pm$0.000 2           \\
        Mean anomaly, $M$ ($^{\circ}$)                         & = & 215.718  $\pm$0.004           & 121.353 65$\pm$0.000 05         
                                                                   & 102.135 8$\pm$0.000 3         &  28.143 7$\pm$0.000 5           \\
        Perihelion, $q$ (au)                                   & = &   0.539 00  $\pm$0.000 03     &   0.864 8304$\pm$0.000 0006   
                                                                   &   0.623 264$\pm$0.000 005     &   0.548 8184$\pm$0.000 0011     \\
        Aphelion, $Q$ (au)                                     & = &   1.459 778 $\pm$0.000 010    &   1.137 705816$\pm$0.000 000004
                                                                   &   1.379 29108$\pm$0.000 00003 &   1.454 3585$\pm$0.000 0002     \\
        Absolute magnitude, $H$ (mag)                          & = &  22.6                         &  21.1               
                                                                   &  21.7                         &  21.7                           \\
       \hline
      \end{tabular}
      \label{elements}
     \end{table*}
%
%

     Results in the figures have been obtained using initial conditions (positions and velocities referred to the barycentre of the Solar 
     system) provided by the Jet Propulsion Laboratory (JPL) HORIZONS system (Giorgini et al. 1996; Standish 1998) and relative to the JD 
     245\,7000.5 epoch which is the $t$ = 0 instant. In addition to the calculations completed using the nominal orbital elements in Table 
     \ref{elements}, we have performed 75 control simulations for each object with sets of orbital elements obtained from the nominal ones 
     within the accepted uncertainties (up to 6$\sigma$) that reflect the observational incertitude in astrometry. In any case, the control 
     orbits start very close to the nominal ones as the Gaussian errors are quite small (see Table \ref{elements}). The computed set of 
     control orbits follows a normal distribution in the six-dimensional orbital parameter space. The orbital evolution is computed in both 
     directions of time at least for 30 kyr. Integration times are longer for the most dynamically stable objects. For clarity, the figures 
     may display just a fraction of the total simulated time. Only a few representative orbits are displayed in the figures.
%
%
     \begin{figure}
       \centering
        \includegraphics[width=\linewidth]{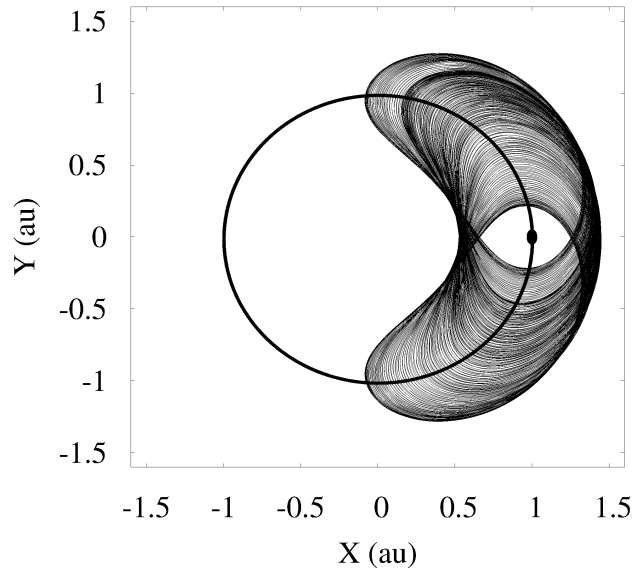}
        \caption{The motion of 2014~OL$_{339}$ over the time range (-150, 150) yr is displayed projected onto the ecliptic plane in a
                 coordinate system rotating with the Earth (nominal orbit in Table \ref{elements}). The orbit and position of our planet are 
                 also indicated. All the investigated control orbits ($\pm6\sigma$) exhibit the same behaviour within this timeframe.
                }
        \label{qs}
     \end{figure}
%
%

  \section{Asteroid 2014~OL$_{339}$, an Aten quasi-satellite}
     Asteroid 2014~OL$_{339}$ was serendipitously discovered by O. Vaduvescu, F. Char, V. Tudor, T. Mocnik, V. Dhillon and D. Sahman 
     observing for EURONEAR (Vaduvescu et al. 2008) from La Palma on 2014 July 29 (Vaduvescu et al. 2014). The object was first detected 
     using the 2.5 m Isaac Newton Telescope at an apparent $R$ magnitude of 21.8. The intended target of the programme was the Apollo 
     asteroid 2013~VQ$_{4}$ but 2014~OL$_{339}$ was visible as a streak near the edge of the observed field. With a value of the semimajor 
     axis, $a$, equal to 0.9994 au, very close to that of our planet (0.9992 au), this Aten asteroid is an NEO moving in an eccentric, $e$ = 
     0.46, and moderately inclined, $i = 10\fdg19$, orbit that makes it an Earth and Venus crosser, and a Mars grazer. Therefore, its orbit 
     is different from those of the three previously known Earth quasi-satellites (see Table \ref{elements}): (164207) 2004~GU$_{9}$, 
     (277810) 2006~FV$_{35}$ and 2013~LX$_{28}$. It is an Aten, not an Apollo, and its eccentricity is the highest of the group which 
     implies that it has the shortest perihelion and the farthest aphelion distances. The source of the Heliocentric Keplerian osculating 
     orbital elements and uncertainties in Table \ref{elements} is the JPL Small-Body Database.\footnote{http://ssd.jpl.nasa.gov/sbdb.cgi}

     Its very small relative semimajor axis, $|a - a_{\rm Earth}| \sim$ 0.000 197$\pm$0.000 007 au (the smallest found so far), makes this 
     object a clear candidate to be an Earth co-orbital. It completes one orbit around the Sun in 364.92 d or 1.00 yr. Its current orbit is 
     based on 27 observations with a data-arc span of 36 d. As expected of a recent discovery, the quality of the orbit of 2014~OL$_{339}$ 
     is at present lower than that of the other minor bodies in Table \ref{elements}. However, it is similar or even better than the one 
     available when the other objects were recognized as unbound companions to our planet. Asteroid 2014~OL$_{339}$ has $H$ = 22.6 mag 
     (assumed $G$ = 0.15) or a diameter of 90 to 200 m for an assumed albedo in the range 0.20--0.04. It is, therefore, smaller than the 
     previously known Earth quasi-satellites (see Table \ref{elements}).

     The motion of 2014~OL$_{339}$ over the time range (-150, 150) yr as seen in a coordinate system rotating with the Earth projected onto 
     the ecliptic plane is plotted in Fig. \ref{qs} (nominal orbit in Table \ref{elements}). This minor body is an Earth co-orbital 
     currently following a quasi-satellite orbit around our planet (see Mikkola et al. 2006; Sidorenko et al. 2014). Due to its significant 
     eccentricity and in accordance to theoretical predictions (Namouni, Christou \& Murray 1999; Namouni \& Murray 2000), the libration 
     angle is rather large. The libration centre corresponds to our planet. Asteroid 2014~OL$_{339}$ appears to pursue a precessing 
     kidney-shaped retrograde path when viewed from our planet over the course of a sidereal year. All the investigated control orbits 
     ($\pm6\sigma$) exhibit the same behaviour within the timeframe mentioned above.
%
%
     \begin{figure*}
       \centering
        \includegraphics[width=\linewidth]{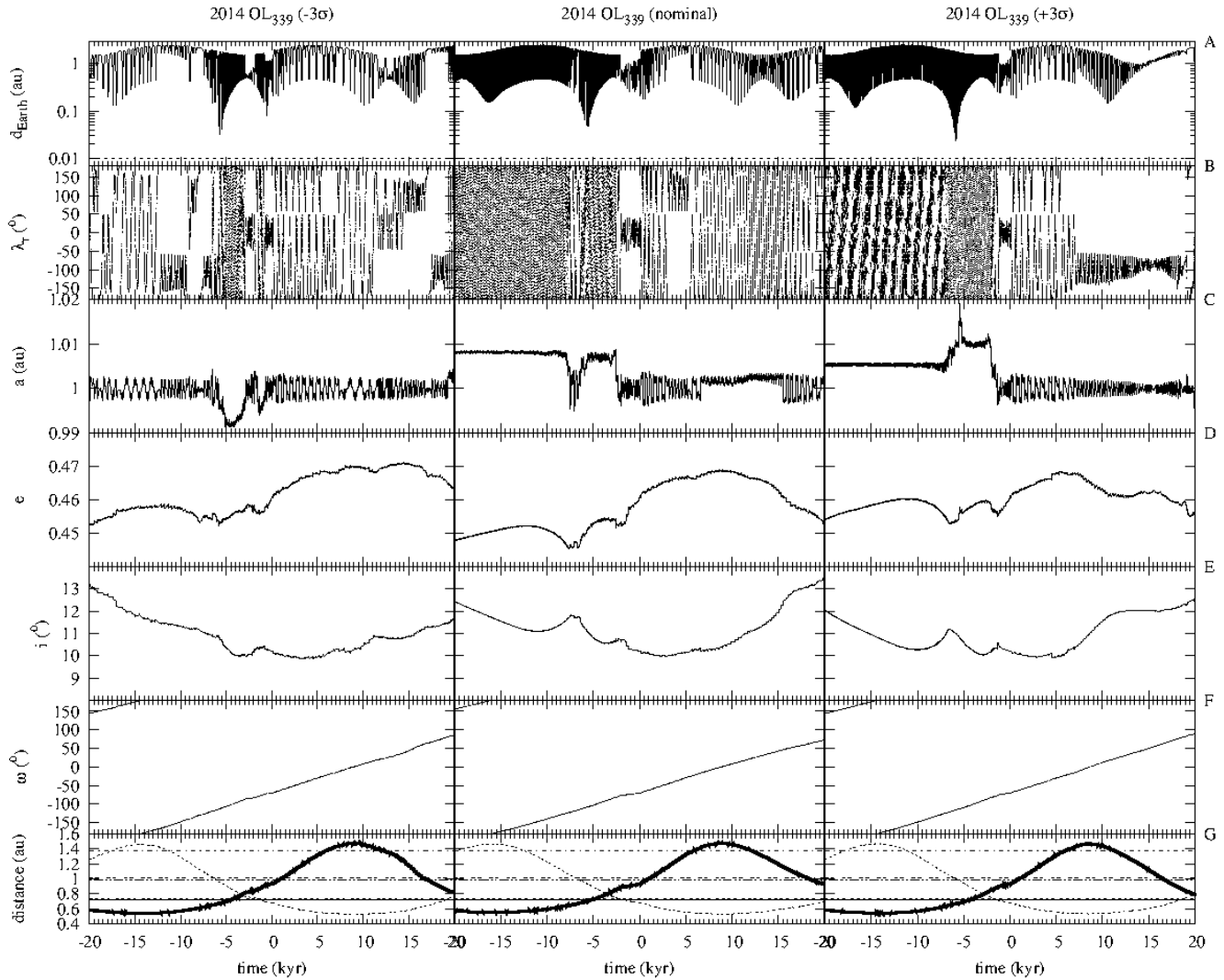}
        \caption{Comparative short-term dynamical evolution of various parameters for an orbit arbitrarily close to the nominal one of
                 2014~OL$_{339}$ as in Table \ref{elements} (central panels) and two representative examples of orbits that are most
                 different from the nominal one (see the text for details). The distance from the Earth (A-panels); the value of the Hill
                 sphere radius of the Earth, 0.0098 au, is displayed. The resonant angle, $\lambda_{\rm r}$ (B-panels). The orbital elements
                 $a$ (C-panels), $e$ (D-panels), $i$ (E-panels) and $\omega$ (F-panels). The distances to the descending (thick line) and 
                 ascending nodes (dotted line) appear in the G-panels. Earth's, Venus' and Mars' aphelion and perihelion distances are also 
                 shown.
                }
        \label{control}
     \end{figure*}
%
%
%
%
     \begin{figure*}
       \centering
        \includegraphics[width=\linewidth]{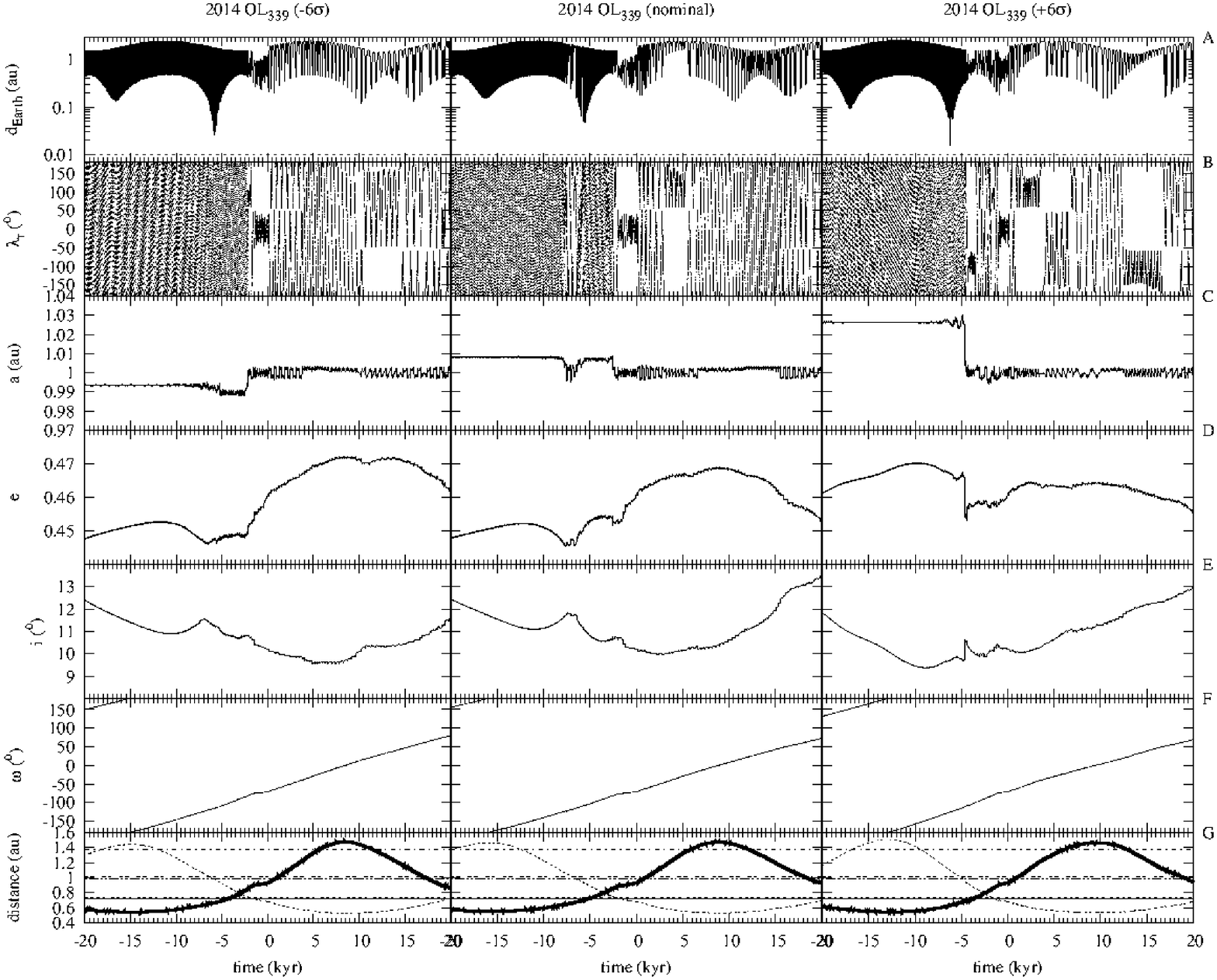}
        \caption{Same as Fig. \ref{control} but for $\pm6\sigma$ deviations.
                }
        \label{control2}
     \end{figure*}
%
%
 
     All the integrated control orbits for 2014~OL$_{339}$ exhibit quasi-satellite libration ($\lambda_{\rm r}$ oscillates around 0\degr) 
     with respect to the Earth at $t$ = 0; the object is a quasi-satellite to our planet at a confidence level $>$ 99.99 per cent (see 
     Figs \ref{control} and \ref{control2}). This co-orbital episode started at least 775 yr ago and it will end 165 yr from now; the 
     duration of the entire quasi-satellite resonance is, in average, approximately 1 kyr (and certainly less than 2.5 kyr), i.e. its 
     current dynamical status is only temporary. The eviction at 165 yr from now coincides with a relatively distant close encounter with 
     our planet at 0.13 au. Prior to the current quasi-satellite episode, the object was probably also co-orbital with our planet, an L$_4$ 
     or L$_5$ Trojan ($\sim$70 per cent) or a horseshoe ($\sim$20 per cent) or, perhaps, a passing object ($\sim$10 per cent) but still in
     the immediate neighbourhood of Earth's co-orbital region. After leaving its current state, it may become an L$_5$ Trojan ($\sim$10 per 
     cent) or, more likely, a horseshoe librator ($\sim$90 per cent). Due to its significant eccentricity and in accordance to theoretical 
     predictions (Namouni, Christou \& Murray 1999; Namouni \& Murray 2000), the libration angle as Trojan is greater than the usual value 
     of $\pm$60\degr, i.e. the libration centre is displaced from the typical equilateral location.

     The overall evolution of all the control orbits within the time interval (-775, 165) yr is virtually identical but beyond those time 
     boundaries, the past and future orbital evolution of this object becomes difficult to predict although it remains in the neighbourhood 
     of Earth's co-orbital region for thousands of years. As an example, Fig. \ref{control} displays the short-term dynamical evolution of 
     an orbit arbitrarily close to the nominal one (central panels) and those of two representative worst orbits which are different from 
     the nominal one. The orbit labelled as `-3$\sigma$' (left-hand panels) has been obtained by subtracting thrice the uncertainty from the 
     orbital parameters (the six elements) in Table \ref{elements}. It has the lowest values of $a$, $e$ and $i$ at the 3$\sigma$ level. In 
     contrast, the orbit labelled as `+3$\sigma$' (right-hand panels) was computed by adding three times the value of the uncertainty to the 
     orbital elements in Table \ref{elements}. This trajectory has the largest values of $a$, $e$ and $i$ (within 3$\sigma$). Asteroid 
     2014~OL$_{339}$ was considerably more stable in the past. It may remain as a co-orbital to our planet switching between the various 
     co-orbital states for many kyr. The values of its semi-major axis (C-panels), eccentricity (D-panels) and inclination (E-panels) remain 
     fairly constant during the entire co-orbital evolution and the object stays well beyond the Hill sphere of our planet (A-panels). The 
     value of its argument of perihelion circulates (F-panels). The results of our calculations show that the true phase-space trajectory 
     followed by this object will diverge exponentially from that obtained from the nominal orbital elements in Table \ref{elements} within 
     a relatively short time-scale; its e-folding time is of the order of 1 kyr. An additional test for consistency is given in Fig. 
     \ref{control2} where the orbital elements have been further modified at the $\pm6\sigma$ level. The short-term dynamical evolution is 
     still consistent with that in Fig. \ref{control} although the object was not co-orbital with our planet a few thousand years into the 
     past. We can certainly state that the probability of this object being a currently active quasi-satellite of our planet is 0.999\,9966.

  \section{Earth quasi-satellites: a review}
     The subject of currently active Earth quasi-satellites has not been revisited recently even if the orbits of those objects recognized 
     as such have been significantly improved in recent times. Here, we provide a brief review of the current dynamical status of (164207) 
     2004~GU$_{9}$, (277810) 2006~FV$_{35}$ and 2013~LX$_{28}$, using their latest orbital solutions (see Table \ref{elements}).

     \subsection{(164207) 2004~GU$_{9}$}
        Asteroid 164207 was discovered by M. Blythe, F. Shelly, M. Bezpalko, R. Huber, L. Manguso, D. Torres, R. Kracke, M. McCleary, H. 
        Stange and A. Milner observing for the Lincoln Near-Earth Asteroid Research (LINEAR) project from Socorro, New Mexico, on 2004 April 
        13 (Kornos et al. 2004) with the 1.0 m LINEAR telescope. At discovery time its apparent magnitude was 19.6. The orbit of this 
        Potentially Hazardous Asteroid (PHA) of the Apollo class has a value of the semimajor axis $a$ = 1.0013 au. Its orbital eccentricity 
        and inclination are moderate, $e$ = 0.14 and $i$ = 13\fdg6. With such an orbit, 164207 always remains in the neighbourhood of the 
        orbit of the Earth--Moon system; no close encounters with other inner planets are possible (see Table \ref{elements}). Its current 
        orbit is based on 175 observations with a data-arc span of 4\,718 d. Besides its orbit, little else is known about 164207: its 
        absolute magnitude has a value of 21.1 mag and its albedo is 0.219 with a diameter of 163 m (Mainzer et al. 2011).
%
%
     \begin{figure*}
       \centering
        \includegraphics[width=\linewidth]{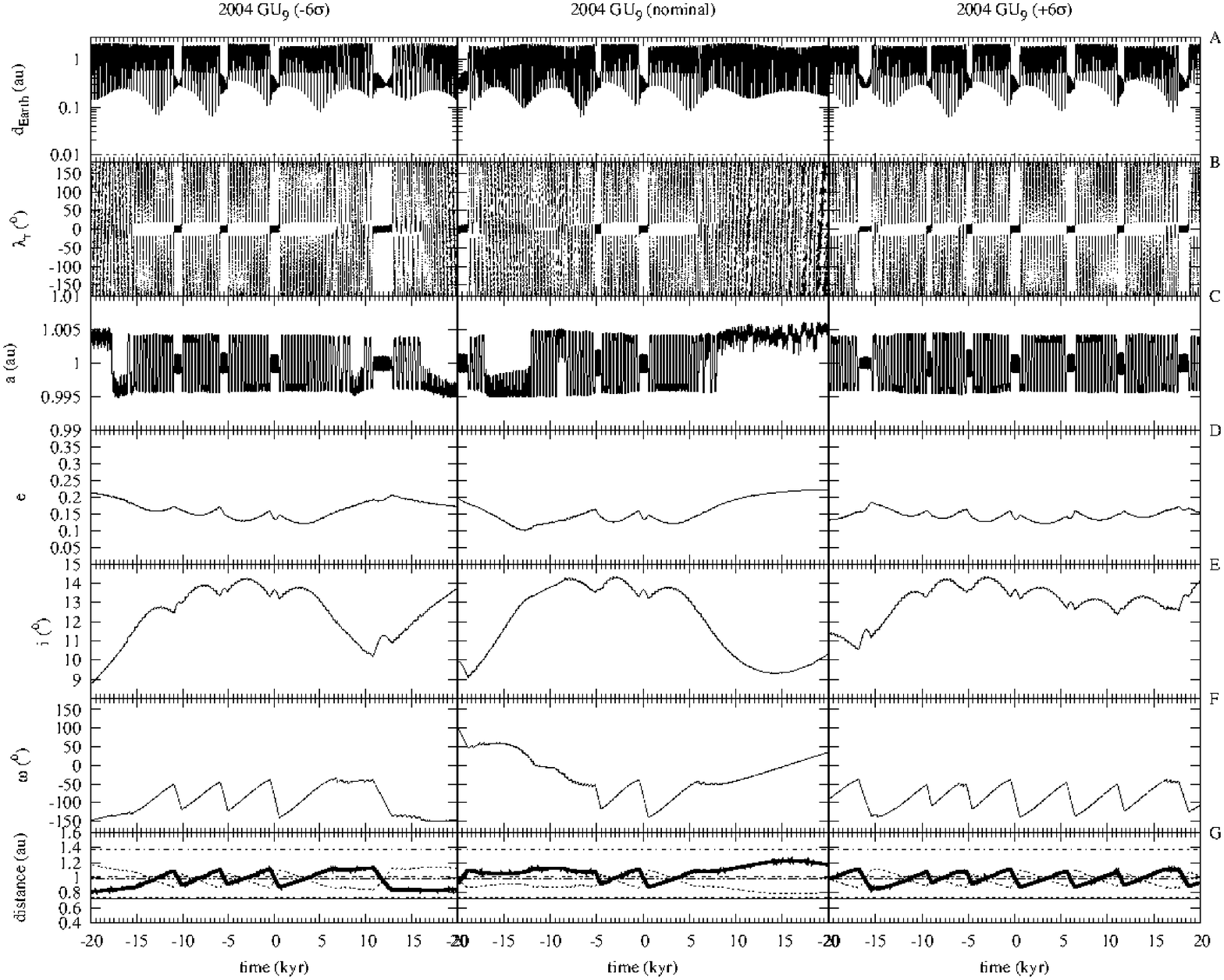}
        \caption{Same as Fig. \ref{control2} but for (164207) 2004~GU$_{9}$.
                }
        \label{gu9}
     \end{figure*}
%
%

        Asteroid 164207 was recognized as a relatively long-lived quasi-satellite companion to the Earth by Connors et al. (2004) and its 
        dynamics was further studied by Mikkola et al. (2006) and Wajer (2010). With the orbit available at the time, these studies 
        concluded that the minor body would remain a quasi-satellite of our planet for several hundred years. Prior to its current dynamical 
        state, 164207 had been a horseshoe librator to the Earth for many thousands of years, its $\lambda_{\rm r}$ oscillating around 
        180\degr. Using the latest orbital solution, all the integrated control orbits for 164207 (within 6$\sigma$) exhibit quasi-satellite 
        libration with respect to the Earth at $t$ = 0 (see Fig. \ref{gu9}). The historical and future evolution of all the control orbits 
        computed coincide in painting an evolutionary track dotted by multiple quasi-satellite resonant episodes of relatively 
        short-duration, just a few kyr or less (see B-panels, Fig. \ref{gu9}). Most of the time, the object is a horseshoe librator to our 
        planet. Transitions between the two resonant states are not triggered by particularly close encounters with the Earth--Moon system 
        but by the persistent action of other mean motion resonances. Asteroid 164207 orbits the Sun in a near 13:8 resonance with Venus so 
        this planet completes 13 orbits around the Sun in the same amount of time the asteroid completes 8. This fact was already pointed 
        out by Wajer (2010). The timings of the transitions depend strongly on the initial conditions. The orbit of this object cannot be 
        predicted with enough certainty beyond a few thousand years. Its present co-orbital episode started about 450 yr ago and it will end 
        nearly 570 yr from now; the duration of the entire quasi-satellite resonance is, in average, nearly 1 kyr with very little 
        dispersion, i.e. like 2014~OL$_{339}$ its current dynamical status is only temporary. Prior to its current engagement as 
        quasi-satellite, this minor body was very probably a horseshoe ($\sim$100 per cent). After leaving its current state, it will return
        to be a horseshoe librator ($\sim$100 per cent). For this object, horseshoe episodes last in average 4 to 6 kyr. These results are
        consistent with those from previous studies.

     \subsection{(277810) 2006~FV$_{35}$}
        Asteroid 277810 was discovered by J. V. Scotti observing from the Steward Observatory at Kitt Peak for the Spacewatch project on 
        2006 March 29 (Gilmore et al. 2006). The object was detected using a 0.9 m telescope at an apparent magnitude of 21.0. With a value 
        of the semimajor axis $a$ = 1.0013 au, this Apollo asteroid is an NEO moving in an eccentric, $e$ = 0.38, and slightly inclined, $i$ 
        = 7\fdg10, orbit that makes it cross the orbits of Venus and the Earth--Moon system (see Table \ref{elements}). Its current orbit is 
        based on 94 observations with a data-arc span of 6\,931 d. Although the object has been observed for almost two decades (the first 
        known pre-discovery observations were made on 1995 April 1), little else besides the orbit is known about 277810; its absolute 
        magnitude, $H$ = 21.7 (assumed $G$ = 0.15), indicates a diameter in the range 130--300 m for an assumed albedo in the range 
        0.20--0.04.
%
%
     \begin{figure*}
       \centering
        \includegraphics[width=\linewidth]{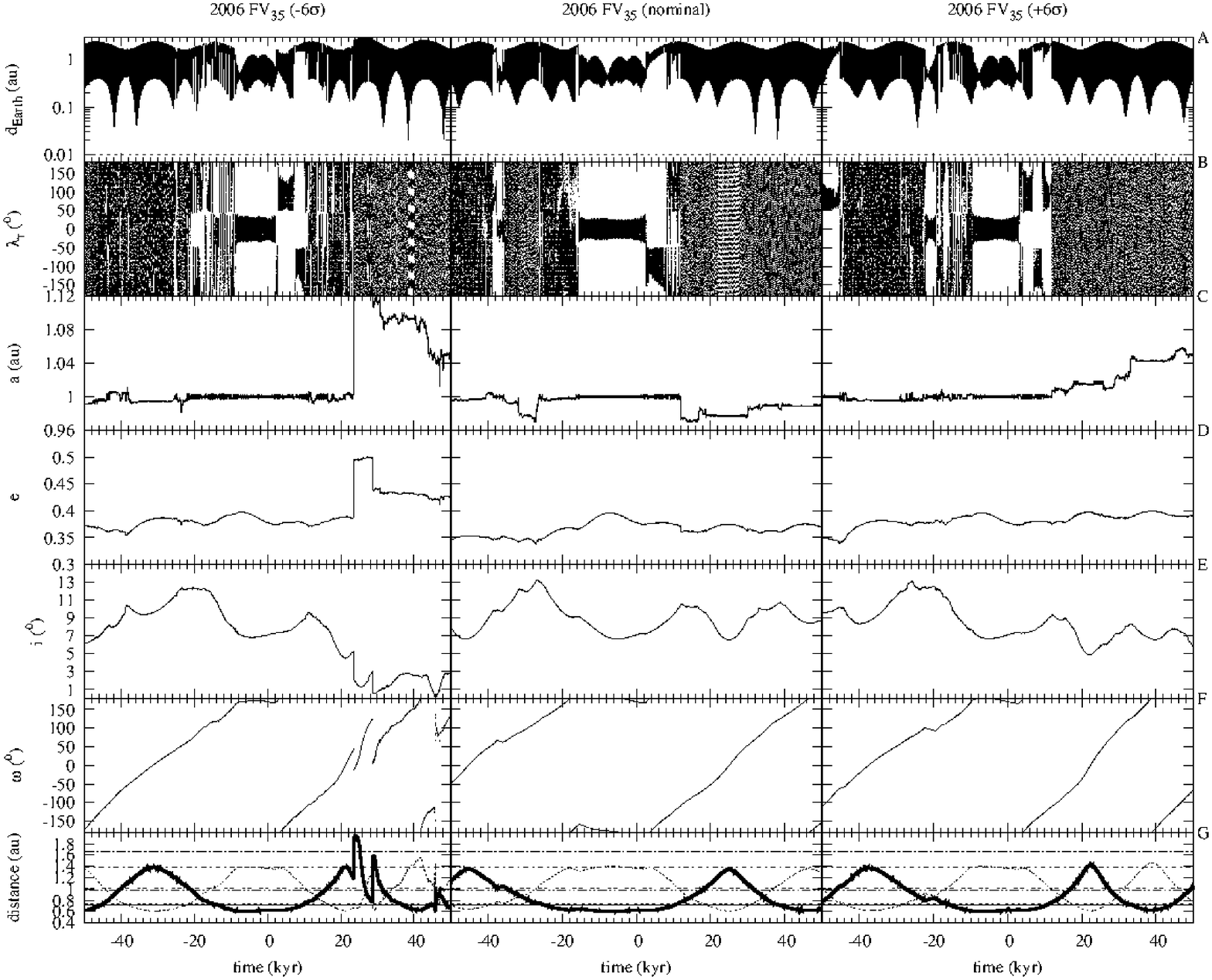}
        \caption{Same as Fig. \ref{control2} but for (277810) 2006~FV$_{35}$.
                }
        \label{fv35}
     \end{figure*}
%
%

        Asteroid 277810 was first reported to be a quasi-satellite of our planet by Wiegert et al. (2008). Its dynamics was further studied 
        by Wajer (2010) who found that it will remain in its present quasi-satellite state for more than 10 kyr. Our calculations (see Fig. 
        \ref{fv35}) confirm that 277810 is experiencing at present a quasi-satellite resonant episode. The object is currently far more 
        stable than 164207. In contrast with the previous object, 277810 rarely follows a horseshoe path and Trojan episodes are far more 
        common. In Fig. \ref{fv35}, B-panels, we observe that the relative mean longitude can librate around 60$^{\circ}$, then the object 
        is called an $L_4$ Trojan, or around -60\degr (or 300\degr), then it is an $L_5$ Trojan. In this case, the timings of the 
        transitions coincide with relatively distant --beyond the Hill radius of our planet (0.0098 au)-- close encounters with the 
        Earth--Moon system. Its current co-orbital episode started at least 8 kyr ago and it will end about 3 kyr from now; the duration of 
        the entire quasi-satellite resonance is, in average, approximately 18 kyr (and certainly less than 22 kyr), i.e. its current 
        dynamical status is still temporary. Prior to the current quasi-satellite episode, the object was probably also co-orbital with our 
        planet, a horseshoe ($\sim$90 per cent) or, perhaps, a passing object ($\sim$10 per cent) but still very close to Earth's co-orbital 
        region. After leaving its current state, it may become a passing object ($\sim$10 per cent) or, more likely, an $L_5$ Trojan 
        ($\sim$90 per cent).    

     \subsection{2013~LX$_{28}$}
        Asteroid 2013 LX$_{28}$ was discovered by N. Primak, A. Schultz, S. Watters and T. Goggia observing for the Pan-STARRS 1 project 
        from Haleakala on 2013 June 12 (Bressi et al. 2013). The object was first observed using a 1.8 m Ritchey-Chretien telescope at an
        apparent magnitude of 20.7. With a value of the semimajor axis $a$ = 1.0016 au, this Apollo asteroid is an NEO moving in a rather 
        eccentric, $e$ = 0.45, and highly inclined, $i$ = 50.0$^{\circ}$, orbit that makes it cross the orbits of Venus and the Earth--Moon 
        system, grazing that of Mars and almost that of Mercury. Its current orbit is based on 26 observations with a data-arc span of 349 
        d. As a recent discovery, little else besides its orbit is known about this object; its absolute magnitude, $H$ = 21.7 (assumed $G$ 
        = 0.15), suggests a diameter in the range 130--300 m for an assumed albedo in the range 0.20--0.04. The object was proposed as a 
        Kozai-resonating Earth quasi-satellite by Connors (2014), who pointed out its remarkable stability. 
%
%
     \begin{figure*}
       \centering
        \includegraphics[width=\linewidth]{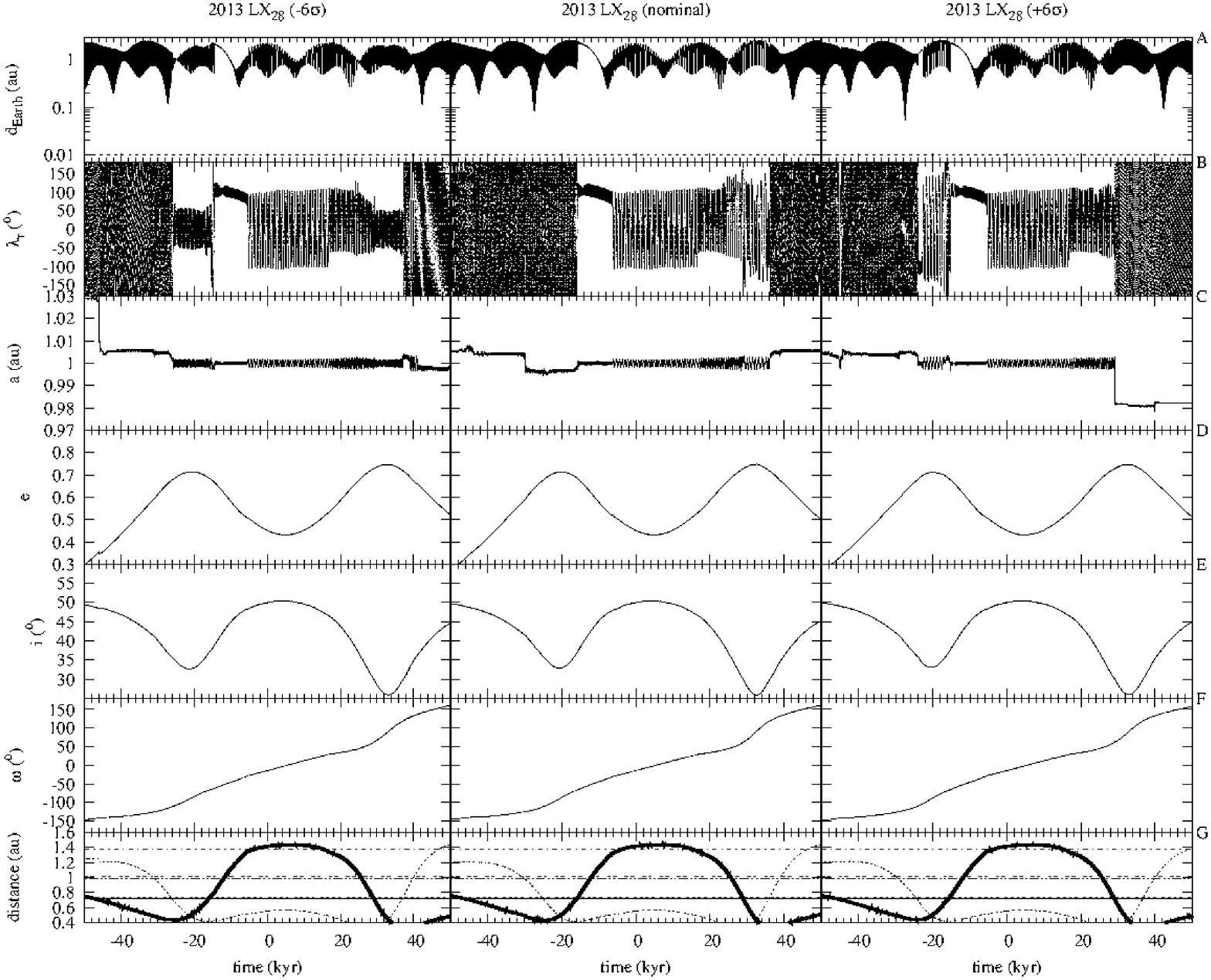}
        \caption{Same as Fig. \ref{control2} but for 2013~LX$_{28}$.
                }
        \label{lx28}
     \end{figure*}
%
%
        
        Once an object is trapped in a 1:1 mean motion resonance and depending on its relative energy with respect to the host planet 
        (H\'enon 1969), it can describe any of the three main orbit types: quasi-satellite, tadpole or horseshoe. Compound states are also 
        possible in which the object may librate around 0$^{\circ}$ with an amplitude $>$ 180$^{\circ}$ encompassing L$_4$ and L$_5$ 
        (compound quasi-satellite-tadpole orbit), asymmetric horseshoe orbits (horseshoe-quasi-satellite orbiters) in which the libration 
        amplitude $>$ 270$^{\circ}$, encompassing the planet, and a few other combinations (see, e.g. Namouni 1999; Namouni et al. 1999). 
        These are typical of objects moving in high-eccentricity, high-inclination orbits and this is what is observed in Fig. \ref{lx28}. 
        Although 2013~LX$_{28}$ is neither a quasi-satellite (see Fig. \ref{lx28}, B-panels) nor a Kozai-resonating body (see Fig. 
        \ref{lx28}, F-panels) in strict sense; $\lambda_{\rm r}$ librates around 0\degr with amplitude $>$ 120\degr and $\omega$ does not 
        librate (or, at least, does not complete a Kozai cycle) just remains relatively close to 0\degr during the entire compound 
        quasi-satellite-tadpole episode. When 2013 LX$_{28}$ leaves the 1:1 mean motion resonance or before entering it, its argument of 
        perihelion is no longer close to 0\degr. This behaviour is fully consistent across the set of simulations. Its current co-orbital 
        episode with our planet began at least 5.5 kyr ago and it will end about 16--30 kyr from now; the duration of the entire compound 
        quasi-satellite resonance is, in average, approximately 35 kyr (and certainly less than 45 kyr), i.e. its current dynamical status 
        is also temporary. However, its compound resonant state changes a few times during that timeframe, the libration amplitude varies 
        although $\lambda_{\rm r}$ still librates around 0\degr. Prior to the current quasi-satellite episode, the object was probably also 
        co-orbital with our planet, an $L_4$ Trojan ($\sim$50 per cent) or, perhaps, a passing object ($\sim$50 per cent) but still very 
        close to Earth's co-orbital region. After leaving its current state, it may become a passing object ($\sim$50 per cent) or an 
        $L_5$ Trojan ($\sim$50 per cent). These Trojan episodes last nearly 10 kyr and are rather asymmetric due to the high-eccentricity, 
        high-inclination orbit. 

  \section{Comparative dynamical evolution of known Earth quasi-satellites}
\begin{figure*}
       \centering
        \includegraphics[width=\linewidth]{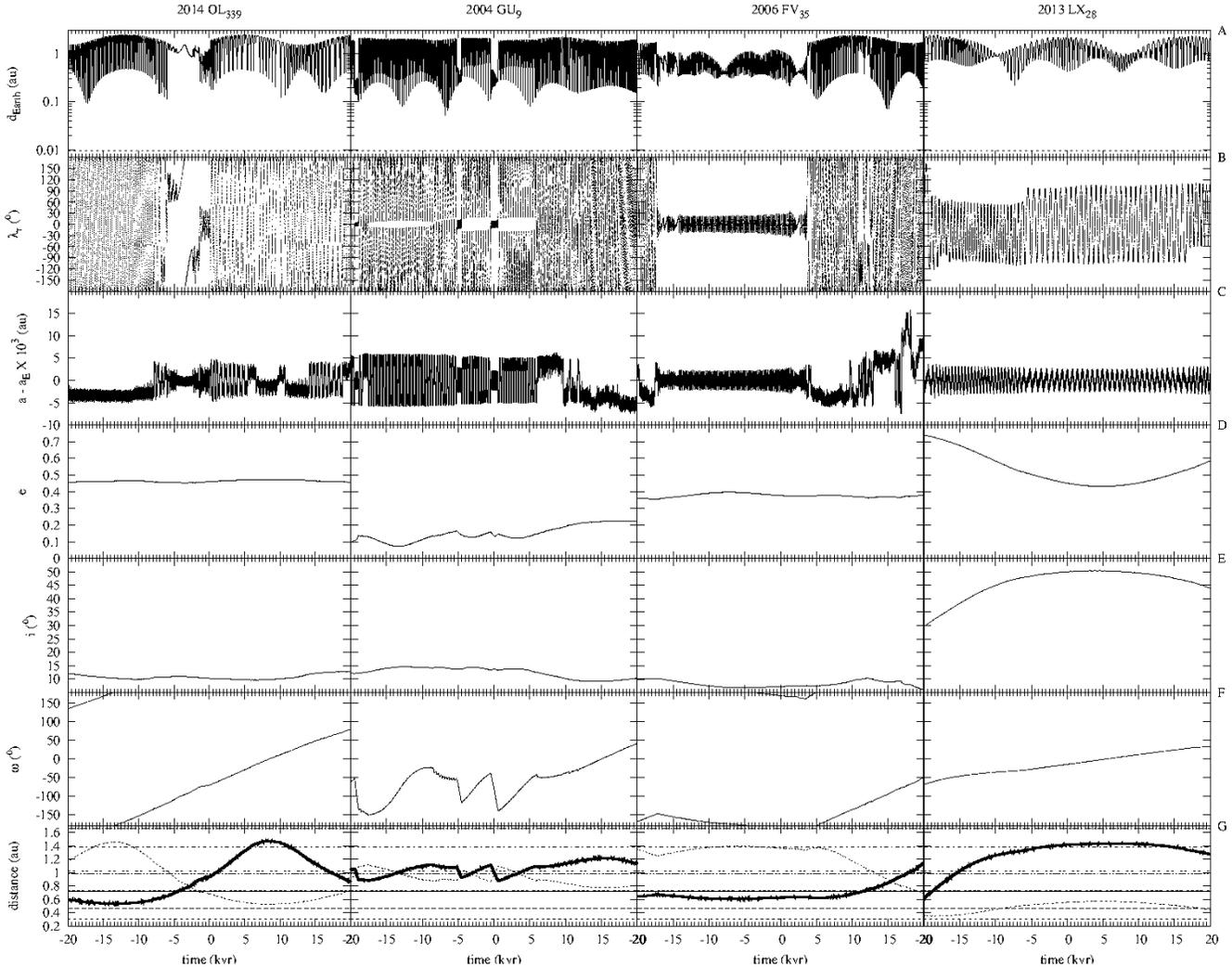}
        \caption{Comparative dynamical evolution of various parameters for the four known Earth quasi-satellites (nominal orbits as in Table 
                 \ref{elements}). The distance from the Earth (panel A); the value of the Hill sphere radius of the Earth, 0.0098 au, is 
                 displayed. The resonant angle, $\lambda_{\rm r}$ (panel B) for the nominal orbit in Table \ref{elements}. The orbital 
                 elements $a - a_{\rm Earth}$ (panel 
                 C), $e$ (panel D), $i$ (panel E) and $\omega$ (panel F). The distances to the descending (thick line) and ascending nodes 
                 (dotted line) appear in panel G. Mars', Earth's, Venus' and Mercury's aphelion and perihelion distances are also shown.
                }
        \label{all}
     \end{figure*}

     Figure \ref{all} displays the comparative evolution of the osculating orbital elements and other parameters of interest of all the
     known Earth quasi-satellites (nominal orbits in Table \ref{elements}). It is clear that this group of objects is, dynamically speaking, 
     very heterogeneous. In particular, three objects exhibit Kozai-like dynamics (see F-panels), see Kozai (1962) and Namouni (1999) for
     technical details: the argument of perihelion of (164207) 2004~GU$_{9}$ oscillates around -90\degr, the one of (277810) 2006~FV$_{35}$ 
     librates around 180\degr, and that of 2013~LX$_{28}$ remains around 0\degr. The argument of perihelion of 2014~OL$_{339}$ circulates. 
     Some Venus co-orbitals (see e.g. de la Fuente Marcos \& de la Fuente Marcos 2013a; de la Fuente Marcos \& de la Fuente Marcos 2014) 
     also exhibit Kozai-like dynamics (see fig. 4, F-panels, in de la Fuente Marcos \& de la Fuente Marcos 2014). In particular, the value 
     of the argument of perihelion of 2002~VE$_{68}$ (also a quasi-satellite) remains close to zero during its entire quasi-satellite
     evolution. For eccentric co-orbitals, this type of resonance provides a temporary effective protection mechanism against close 
     encounters with the host planet: the Earth for 2013~LX$_{28}$ and Venus for 2002~VE$_{68}$. In this case, the nodes are located at 
     perihelion and at aphelion, i.e. away from the host planet (see e.g. Milani et al.  1989).

     Asteroid 2014~OL$_{339}$ is an Aten, the other three confirmed quasi-satellites are Apollos although the reason for the absence of the 
     Kozai resonance in the case of 2014~OL$_{339}$ is not this but its relatively large eccentricity. For an object following an inclined 
     path, close encounters with major planets are only possible in the vicinity of the nodes. The distance between the Sun and the nodes is 
     given by $r = a (1 - e^2) / (1 \pm e \cos \omega)$, where the `+' sign is for the ascending node and the `-' sign is for the descending 
     node. The positions of the nodes are plotted in the G-panels of Fig. \ref{all}. The descending node of 2014~OL$_{339}$ is close to the 
     orbit of the Earth, its ascending node is near Venus. In contrast, both nodes of 164207 are currently near the Earth, the ascending 
     node of 277810 is perturbed by Mars and the descending one is relatively free from perturbations by Venus; the descending node of 
     2013~LX$_{28}$ is perturbed by Mars and its ascending one is also relatively free from perturbations by Venus. The Kozai resonance is 
     effective in protecting the paths of 277810 and 2013~LX$_{28}$ against close encounters with the Earth--Moon system as their nodes are 
     away from it, stabilizing their orbits but makes the orbit of 164207 rather unstable. Here, the libration occurs at $\omega$ = -90\degr, not 
     0\degr or 180\degr. Under these circumstances, aphelion and perihelion always occur away from the ecliptic plane. A common feature of 
     the orbital evolutions of 164207 and 2014~OL$_{339}$ is in their enhanced instability when compared to the other two. This translates 
     into relatively frequent episodes in which we observe switching between resonant states. Transfers between tadpole, horseshoe and 
     quasi-satellite orbits are triggered by close encounters with the inner planets and those are the result of the libration of the nodes 
     (Wiegert, Innanen \& Mikkola 1998). Asteroids 277810 and 2013~LX$_{28}$ do not exhibit Kozai-like dynamics outside the timeframe in 
     which they are quasi-satellites.

     Although there are no two Earth quasi-satellites alike, the closest dynamical relative to 2014~OL$_{339}$ is 164207. It also stays as
     an Earth quasi-satellite for about 1 kyr (see Fig. \ref{all}, second column, panel G) which is consistent with previous results 
     presented by Wajer (2010). It was a horseshoe librator prior to its capture as quasi-satellite and it will return to that resonant 
     state after its eviction. Figure \ref{all} shows that only the Earth-Moon system plays a significant role in destabilizing its orbit; 
     contrary to previous results in Wajer (2010), Venus does not appear to play a significant role in the current dynamical evolution of 
     this object. The orbits of 164207 and 2014~OL$_{339}$ can only be accurately calculated for a few hundred years forward and backward in 
     time. In sharp contrast, 277810 remains in the quasi-satellite state for a long period of time. Our calculations agree reasonably well 
     with those of Wajer (2010), the object has remained in its current state for more than 15 kyr and it will remain there for a few 
     thousand more years. Discrepancies with Wajer (2010) could be the result of using updated orbits and different physical models. Chaotic 
     orbits are not only sensitive to changes in the initial conditions but also to different dynamical models. Although Connors (2014) 
     classifies 2013~LX$_{28}$ as quasi-satellite, this is incorrect in strict sense because its orbit is hybrid. It is a persistent 
     co-orbital companion to the Earth that follows a compound quasi-satellite-tadpole orbit that encloses Earth's Lagrangian points L$_5$ 
     and L$_4$, as well as the Earth itself (see e.g. Namouni 1999; Namouni et al. 1999). In principle, close encounters are possible with 
     Mercury, Venus, the Earth and Mars but the asteroid is temporarily protected against close approaches by a Kozai-like resonance with 
     Jupiter. Its dynamics is somewhat similar to that of the well studied Apollo asteroid 10563 Izhdubar (1993 WD) although the argument of 
     perihelion of this object librates around 90\degr (see Christou 2000a) not 0\degr. Even if not strictly a quasi-satellite, 
     2013~LX$_{28}$ is the most stable of the group with a most probable duration of its current state in the range of 35 to 45 kyr.
%
%
%
%

  \section{Conclusions}
     In this paper, we have identified yet another Earth quasi-satellite, 2014~OL$_{339}$. Its dynamical status is temporary and it is not 
     expected to last more than 1 to 2 kyr as this object is one of the most unstable known Earth quasi-satellites; its e-folding time is
     $\sim$1 kyr. In the Solar system and among the terrestrial planets, the Earth has the largest number of detected quasi-satellites with
     four; this is likely to be the result of observational bias, though. A comparative analysis of the short-term dynamical evolution of 
     these objects shows that they are, dynamically speaking, very heterogeneous although three objects exhibit Kozai-like dynamics. The 
     identification of Kozai librators among members of the NEO population is not new (see e.g. Michel \& Thomas 1996; de la Fuente Marcos 
     \& de la Fuente Marcos 2013b). This indicates that the Kozai resonance plays a significant role in the orbital evolution of many Earth 
     quasi-satellites and also in the chain of events that drives them into this particular resonance and away from it. In this context, 
     2014~OL$_{339}$ is an outlier as it is the only currently known Earth quasi-satellite not to be submitted to a Kozai resonance. Given
     their current orbits, none of the objects discussed here may impact our planet within the next few hundred --(164207) 2004~GU$_{9}$ 
     and 2014~OL$_{339}$-- or even several thousand --(277810) 2006~FV$_{35}$ and 2013~LX$_{28}$--  years. Although these four objects
     are currently experiencing quasi-satellite episodes within the 1:1 mean motion resonance with the Earth, their dynamical contexts are
     quite different hinting at a richer picture of the quasi-satellite state than conventionally portrayed, with multiple pathways to the
     same resonant phase. The diverse dynamical histories found for the members of this group make a common origin for any pair of them 
     rather unlikely. 

     In this work, relativistic terms and the role of the Yarkovsky and Yarkovsky--O'Keefe--Radzievskii--Paddack (YORP) effects (see e.g.
     Bottke et al. 2006) have been ignored. The non-inclusion of these effects has no impact on the evaluation of the present dynamical
     status of the minor bodies studied here but may affect predictions regarding their future evolution and dynamical history. In 
     particular, the Yarkovsky effect may have a role on the medium- and long-term evolution of objects as small as the minor bodies 
     discussed here. Proper modelling of the Yarkovsky force requires knowledge on the physical properties of the objects involved (for 
     example, rotation rate, albedo, bulk density, surface conductivity, emissivity) which is not the case for these minor bodies. 
     Perturbational effects arising from the co-orbital evolution with our planet may render these non-gravitational effects negligible,
     though.

  \section*{Acknowledgements}
     We would like to thank the referee for his/her prompt, to-the-point report and S. J. Aarseth for providing the code used in this 
     research. This work was partially supported by the Spanish `Comunidad de Madrid' under grant CAM S2009/ESP-1496. We thank M. J. 
     Fern\'andez-Figueroa, M. Rego Fern\'andez and the Department of Astrophysics of the Universidad Complutense de Madrid (UCM) for 
     providing computing facilities. Part of the calculations and the data analysis were completed on the `Servidor Central de C\'alculo' of 
     the UCM and we thank S. Cano Als\'ua for his help during this stage. In preparation of this paper, we made use of the NASA 
     Astrophysics Data System, the ASTRO-PH e-print server, the MPC data server and the NEODyS information service.

\end{document}